\documentclass[12pt]{iopart}

\usepackage{graphicx}
\usepackage{subfigure}

\begin{document}

\title{Dynamics of Competing Ideas in Complex Social Systems}

\author{Yubo Wang$^1$,~Gaoxi Xiao$^1$,~Jian Liu$^2$}
\address{$^1$School of Electrical and Electronic Engineering, Nanyang Technological University, Singapore 639798}
\address{$^2$JPT Electronics Co., Ltd., Shenzhen, P. R. China 518110}
\ead{egxxiao@ntu.edu.sg}

\begin{abstract}
Individuals accepting an idea may intentionally or unintentionally impose influences in a certain neighborhood area, making other individuals within the area less likely or even impossible to accept other competing ideas. Depending on whether such influences strictly prohibit neighborhood individuals from accepting other ideas or not, we classify them into exclusive and non-exclusive influences, respectively. Our study reveals for the first time the rich and complex dynamics of two competing ideas with neighborhood influences in scale-free social networks: depending on whether they have exclusive or non-exclusive influences, the final state varies from multiple coexistence to founder control to exclusion, with different sizes of population accepting each of the ideas respectively.  Such results provide insights helpful for better understanding the spread (and the control of spread) of ideas in human society.
\end{abstract}

\submitto{\NJP}
\maketitle

\section{\label{sec-1}Introduction}
Ideas spread in human society through education, public media, religious practices, literature publications, propaganda and rumors etc. While some ideas can easily spread out with virtually no resistance (e.g., the education of fundamental science in primary schools), others may have to be in face of competitions. The competitions can be mild or even hardly noticeable such as those between different opinions in rumor spreading, or rather fierce, e.g., some violent conflicts between different religions in human history. 

While the spreading of an idea with no competitors, which to a certain extent is analogous to the spread of an infectious disease, has been extensively discussed \cite{ref-1}, studies on dynamics of competing ideas are largely in absence. In fact, even the existing work on spreading of multiple competing viruses/pathogens is very limited. The majority part of the existing work is on analyzing competing viruses in well-mixed populations (e.g., \cite{ref-2,ref-3,ref-4,ref-5,ref-6,ref-7}), with no detailed modeling of the interactions between individuals. It is only in recent years that a few detailed studies have been conducted on interacting viruses  with the aid of  graph theory, considering (i) \textit{cross protection} where individuals infected by one agent are immunized to the other \cite{ref-8,ref-9}; (ii) propagations of two agents in two overlay networks \cite{ref-10}; and (iii) a special case where agent A induces agent B which in turn suppresses agent A \cite{ref-11}, respectively. In social science, various voter models have been proposed for studying the dynamics of two different opinions. Typically it is assumed that each voter may discard his own opinion and accept one of his randomly selected neighbors' opinion instead (e.g., \cite{ref-12}). Such models help explain the coexistence of different opinions. Yet the assumption that each individual has to accept one of the two opinions at any single moment (S/he cannot be left idle.) makes such models quite specific for studying voter behaviors only. 

We argue that the spreading of competing ideas is very different from the spreading of competing viruses. An important feature of idea spreading is that an idea typically can generate some ``influences'' in a certain neighborhood area. Individuals in the area may not necessarily accept the idea; yet while under the influences, the chance that they accept a different competing idea is usually lowered, or even eliminated in some extreme cases. Such a feature does not exist in most virus spreading cases, and to the best of our knowledge, has never been systematically studied in existing sociology research either. 

In this paper, we focus on studying the effects of such neighborhood influences. Specifically, we consider two representative types of neighborhood influences as follows: 
\begin{itemize}

\item  Knowing that a close friend has accepted an idea may not immediately or finally make us accept the same idea. However, it usually lowers the chance that we accept a different idea, at least within a certain period of time \cite{ref-13,ref-14}. Since the influence from the friend in this example does not eliminate the possibility that we accept a different idea, we term it as \textit{non-exclusive} \textit{influence}. An interesting observation is that when under non-exclusive influence, people sometimes may finally accept multiple different ideas, say, by taking them as valid and valuable insights from different points of view.  

\item  In a region ruled by extremists, people may be prohibited from accepting any other idea, or be deprived of access to any competing ideas altogether \cite{ref-13}. When such control is strictly implemented, the chance that people accept a different idea may be virtually zero. We term such cases as with \textit{exclusive influence}.

\end{itemize}

To evaluate the effects of exclusive and non-exclusive influences in idea spreading, we consider three different cases with two competing ideas where (i) both ideas have non-exclusive influences; (ii) both have exclusive influences; and (iii) the two ideas have non-exclusive and exclusive influences respectively. Considering that many social networks closely resemble scale-free networks with power-law nodal degree distribution \cite{ref-15,ref-16,ref-17}, we focus on studying the spreading of two competitive ideas in scale-free networks.  

For the spread of two competing agents with cross -protection in scale-free networks, it is easy to figure out that the two agents can always coexist in the steady state (though a strict proof has never been published in any reference to the best of our knowledge). Specifically, when two competing agents spread out in scale-free networks following the susceptible-infected-susceptible (SIS) scheme \cite{ref-18,ref-19,ref-20}, at any single moment neither of them can infect all the high-degree hub nodes, unless we assume that at least one of them has nearly infinite transmissibility. It is known that in sufficiently large scale-free networks, leaving a non-zero percentage of hub nodes unprotected causes persistent existence of infection \cite{ref-21}. Therefore the two agents definitely coexist in the steady state. For two competing ideas with exclusive and/or non-exclusive neighborhood influences, however, the dynamics is much richer. Specifically, the main conclusions of our study can be summarized as follows:

\begin{itemize}
\item  For competing ideas both with non-exclusive influences, they may have \textit{multiple coexistence} states: the final states of the two ideas with comparable transmissibility and strong neighborhood influences are determined by their initial densities; while the idea with a relatively higher transmissibility can easily suppress its competitor to a low level.

\item  For two ideas both with exclusive influences, they can \textit{never} stably coexist in scale-free networks regardless of their respective transmissibility. The possible outcomes can be classified into \textit{founder control} \cite{ref-22}, where the final winner is determined by the initial densities of the two ideas, or \textit{exclusion }where one idea steadily drives out the other.

\item  For two ideas with non-exclusive and exclusive influences respectively, the one with exclusive influence has a chance to drive out its competitor altogether. However, this is guaranteed to happen only if its transmissibility is high enough compared to that of its competitor. Since it typically takes nontrivial efforts (energy) to have exclusive influence in a neighborhood region, which may consequently lead to a lower transmissibility, it may \textit{not} be a favorable strategy to try to have exclusive influence. In fact, for both the cases of non-exclusive influence vs. non-exclusive influence and non-exclusive influence vs. exclusive influence, when subject to limited resources, it helps enlarge the size of acceptance at steady state by focusing on increasing the transmissibility of the idea rather than weakening the neighborhood influence of the competitor. 

\end{itemize}

Theoretical analysis and numerical simulations verify the above conclusions in random scale-free networks.

\section{\label{sec-2}Definitions of the models}
The model is defined as follows. Two competing ideas, hereafter termed as idea-I and idea-II respectively, propagate in a scale-free network following the standard SIS epidemiological scheme. We term idea-I as the idea-II's \textit{competitor idea}, and vice versa. For convenience, an individual accepting an idea is termed as being \textit{infected }by the idea; and \textit{susceptible} otherwise. A susceptible individual adjacent to one or more infected individuals is termed as being \textit{exposed} to the idea. In a discrete unity time slot, the two ideas have infection probabilities of $\nu _{1}$ and $\nu _{2}$ respectively on those individuals exposed to only one of them. For individuals exposed to both of them, the infection probabilities become $\alpha \nu _{1} $ for idea-I and $\beta \nu _{2} $ for idea-II, where $\alpha$ and $\beta$ are \textit{influential factors} of idea-I and idea-II respectively, $0 \le \alpha , \beta \le 1$. Obviously, a smaller value of the influential factor corresponds to a stronger suppressing influence imposed by the competitor idea and a zero influential factor denotes that the competitor idea has exclusive influence. Assume that individuals infected by idea-I and idea-II are cured and become susceptible again at probabilities of $\delta _{1} $ and $\delta _{2} $ respectively in unity time. The \textit{spreading rates} of ideas-I and idea-II can therefore be defined as $\lambda _{1} =\nu _{1} /\delta _{1} $ and $\lambda _{2} =\nu _{2} /\delta _{2} $ respectively \cite{ref-20}. 

Note that in the proposed model, we assume that the chance of getting infected depends on the \textit{presence} of infectious neighbors rather than the number of them. A different model can be proposed by assuming that each infected neighbor has an independent chance of transmitting the idea. In fact, both models have been extensively utilized in studies on virus and contagion spreading \cite{ref-20,ref-23,ref-24,ref-25} and they have always led to basically the same conclusions, with (at most) only some differences quantitatively. In this paper, we adopt the former model to allow simpler mean-field analysis. 

To the best of our knowledge, the only existing work which may be regarded as loosely related to the general model above was reported in \cite{ref-26}. The model there was on the spreading of two interacting rumors, one of which is always preferably adopted. It may be viewed as a special case of the proposed model where $\alpha =1$ and $\beta =1-\lambda _{1} $, though in \cite{ref-26} it was assumed that the two rumors can never co-infect an individual and it mainly studied the effects of network structures on coexistence of the rumors rather than dynamics of  rumors with neighborhood influences. 

\section{\label{sec-3}Dynamics of two competing ideas with non-exclusive influences}
\subsection{\label{subsec-3-1}Coexistence of the two ideas}
By adopting the mean-field theory \cite{ref-27}, the spreading of two competing ideas in a random scale-free network can be analyzed. Specifically, we assume that the population is of a fixed unit size  and it can be modeled into a random scale-free network with degree distribution $P(k)\sim k^{-r}$, where nodes represent individuals and links represent possible channels for idea spreading between adjacent individuals, and $P(k)$ the probability that an randomly selected node has $k$ neighbors. Term the densities of $k$-degree nodes infected by idea-I and idea-II at time $t$ as $\rho _{1,k} (t)$ and $\rho _{2,k} (t)$, respectively. Since the instantaneous changing rate of the infection density by an idea equals the density of new infection minus the density of recovery, the time-evolution dynamics of the two ideas can be described by the following coupled equations:
\numparts 
\begin{eqnarray}
\frac{d\rho _{1,k} (t)}{dt} =&  -  \rho _{1,k} (t) \delta_{1} + \left ( 1-\rho _{1,k} (t) \right ) \left [ 1-\left (1-\theta _{1} (t) \right )^{k} \right ]  \nonumber \\ & \times \left ( \left (1-\theta _{2} (t) \right)^{k} \nu_{1} +  \left [1-\left (1-\theta _{2} (t) \right )^{k} \right ] \alpha \nu_{1} \right ), \label{eq-1a} \\
\frac{d\rho _{2,k} (t)}{dt} =&  -\rho _{2,k} (t) \delta_{2} + \left (1-\rho _{2,k} (t) \right ) \left [1-\left (1-\theta _{2} (t)\right )^{k} \right ] \nonumber \\ & \times \left ( (1-\theta _{1} (t))^{k} \nu_{2} + \left [ 1-\left (1-\theta _{1} (t) \right )^{k} \right ] \beta  \nu_{2} \right ), \label{eq-1b}
\end{eqnarray} 
\endnumparts
where  $\theta _{1} (t)$ denotes the probability that a randomly chosen link is connected to an individual infected by idea-I and consequently, $(1-\theta _{1} (t))^{k} $ the probability that a $k$-degree node is not directly connected to any node infected by idea-I. $\theta _{2} (t)$ and $(1-\theta _{2} (t))^{k} $ are defined for idea-II similarly. Note that as pointed out earlier, we allow co-infection of two ideas on the same individual.

In a random scale-free network, the probability that a randomly chosen link points to a $k$-degree node equals $kP(k)/ \langle k \rangle $, where $ \langle k \rangle =\sum _{k}kP(k) $ is the average nodal degree \cite{ref-28}. Therefore $\theta _{1} (t)$ and $\theta _{2} (t)$ can be expressed as
\numparts 
\begin{eqnarray} 
\theta _{1} (t) &= \frac{1}{\langle k \rangle}  \sum _{k}kP(k)\rho _{1,k} (t) , \label{eq-2a} \\
\theta _{2} (t) &= \frac{1}{\langle k \rangle}  \sum _{k}kP(k)\rho _{2,k} (t) . \label{eq-2b}
\end{eqnarray} 
\endnumparts
When the spread of the two ideas reaches the stationary state at time $t\to \infty$, $d\rho_{1,k} /dt = d\rho_{2,k} /dt=0$ in equations~(\ref{eq-1a})-(\ref{eq-1b}) and $d\theta _{1} /dt=d\theta _{2} /dt=0$ in equations (\ref{eq-2a})-(\ref{eq-2b}).  Therefore $\theta _{1} $ and $\theta _{2}$ satisfy
\numparts 
\begin{eqnarray}
\fl \theta _{1}  =  f(\theta _{1} ,\theta _{2} ,\alpha ,\lambda _{1} ) =\frac{1}{\langle k \rangle}  \sum _{k} kP(k) \frac{  \lambda _{1} \left [1-(1-\theta _{1} )^{k} \right ] \left [ \alpha +(1-\alpha )(1-\theta _{2} )^{k} \right ] }{\displaystyle 1+\lambda _{1} \left [ 1-(1-\theta _{1} )^{k} \right ] \left [ \alpha +(1-\alpha )(1-\theta _{2} )^{k} \right ] }, \label{eq-3a} \\
\fl \theta _{2} = f(\theta _{2} ,\theta _{1} ,\beta ,\lambda _{2} ) = \frac{1}{\langle k \rangle}  \sum _{k} kP(k) \frac{ \lambda _{2} \left [ 1-(1-\theta _{2} )^{k} \right ] \left [ \beta +(1-\beta )(1-\theta _{1} )^{k} \right ] }{\displaystyle 1+\lambda _{2} \left [ 1-(1-\theta _{2} )^{k} \right ] \left [ \beta +(1-\beta )(1-\theta _{1} )^{k} \right ] }. \label{eq-3b}
\end{eqnarray} 
\endnumparts
where $\lambda _{1} = \nu _{1} /\delta _{1} $ and $\lambda _{2} = \nu _{2} /\delta _{2} $ are spreading rates of the two ideas as defined in Section \ref{sec-2}. Examining the solutions of $\theta _{1} $ and $\theta _{2} $ yielded from equations (\ref{eq-3a})-(\ref{eq-3b}), the final states of the two ideas can be predicted. It is easy to see that $\theta _{1} =\theta _{2} =0$ is always a solution of equations (\ref{eq-3a})-(\ref{eq-3b}). To have a non-zero solution of $\theta _{1} $, the following inequality 
\begin{equation} \label{eq-4} 
\left. \frac{d}{d\theta _{1} } f(\theta _{1} ,\theta _{2} ,\alpha ,\lambda _{1} )\right|_{\theta _{1} =0} >1 
\end{equation} 
must be satisfied \cite{ref-29}. Bringing the detailed expression of function $f(\theta _{1} ,\theta _{2} ,\alpha ,\lambda _{1} )$ in equation (\ref{eq-3a}) into equation (\ref{eq-4}) and letting $\theta _{1} $ equal 0, we have that the spreading rate of idea-I has to fulfill
\begin{equation} \label{eq-5} 
\lambda _{1} > \frac{\langle k \rangle}{\displaystyle \sum _{k} k^{2} P(k) \left [ \alpha +(1-\alpha )(1-\theta _{2} )^{k} \right ] } , 
\end{equation}
where $\theta _{2}$ is the final state of idea-II. Equation (\ref{eq-5}) reveals that in scale-free networks with exponent $r \le 3$, idea-I will persistently exist (i.e., $\theta _{1} >0$) as $\alpha \sum _{k}k^{2} P(k) $ always goes to infinity for any $\alpha >0$ \cite{ref-20,ref-21,ref-24,ref-29} (The special case where $\alpha =0$ will be analyzed in Sections \ref{sec-4} and \ref{sec-5}.). Similarly, we can derive that idea-II also persistently exists in scale-free networks. The coexistence of non-exclusive competing ideas in scale-free networks therefore can be verified. The detailed dynamics of the two ideas, however, can be rather complex. Later we shall prove that the competing ideas may have multiple coexistences.

\begin{figure}
\begin{center}
\subfigure[]{\label{fig_1a}
\includegraphics[width=50mm,height=50mm]{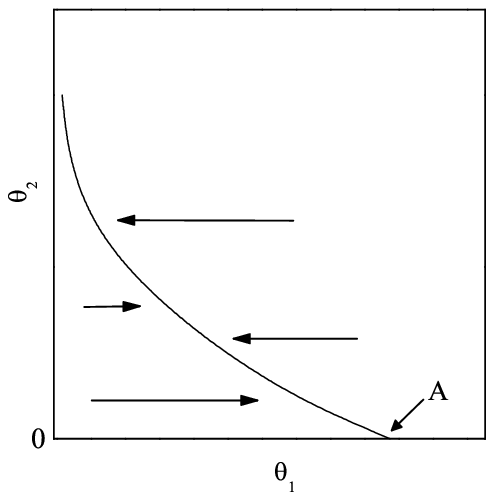}}
\subfigure[]{\label{fig_1b}
\includegraphics[width=50mm,height=50mm]{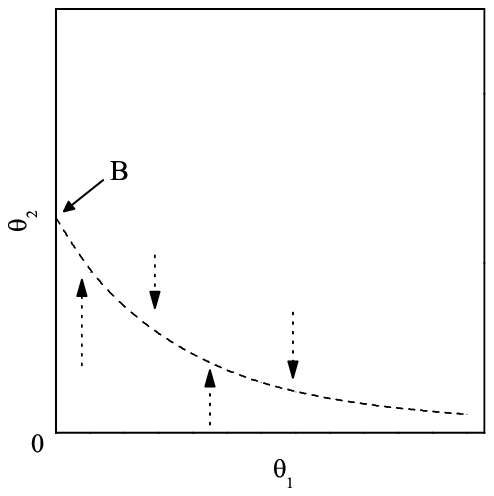}}
\end{center}
\caption{\label{fig_1}Schematic phase plane diagrams of the two ideas' zero-growth isoclines in an infinite scale-free network with exponent $r=3$ and the minimum nodal degree $2$. The spreading rates and influential factors are set to be $\lambda _{1} =0.3$, $\lambda _{2} =0.25$ and $\alpha =0.15$, $\beta =0.35$, respectively.  Arrows indicate the moving directions of $\theta _{1} $ ($\theta _{2}$) when $\theta _{2} $ ($\theta _{1}$) holds as a constant.}
\end{figure}

The two equations~(\ref{eq-3a})-({\ref{eq-3b}}) can be plotted as two separate function curves of $\theta _{1} $ and $\theta _{2} $(known as \textit{zero-growth isoclines} of ideas or in short \textit{isoclines} since they represent the values of $\theta _{1} (t)$ and $\theta _{2} (t)$ at stationary state with a zero growth rate, i.e., where $d\theta _{1} /dt=d\theta _{2} /dt=0$ \cite{ref-22,ref-30}). With the assistance of isoclines we can predict the steady states and phase transition of the idea spreading. An example case is shown in figure 1, where the spreading rates and influential factors of the two ideas are $\lambda _{1} =0.3$, $\lambda _{2} =0.25$ and $\alpha =0.15$, $\beta =0.35$, respectively. Figure~\ref{fig_1a} is for idea-I as described in equation (3a) and figure~\ref{fig_1b} for idea-II as described in equation~(\ref{eq-3b}). More specifically, the isocline in figure~\ref{fig_1a} represents the values of $\theta _{1} $ at steady state where $d\theta _{1} /dt=0$ and $\theta _{2} $ holds as a constant. For any given value of $\theta _{1} $ which is not on the solid line, it moves horizontally towards the solid line as shown in figure~\ref{fig_1a}; otherwise, it stays still. Arrows in figure~\ref{fig_1a} indicate the moving directions of $\theta _{1} $ if it is not on the curve (assuming that $\theta _{2} $ remains as a constant). The intersection point $A$ indicates the value of $\theta _{1} $ in absence of idea-II in the network. As we have proved earlier, equation~(\ref{eq-3a}) always has a positive solution of $\theta _{1} $ for any $\alpha >0$ regardless the value of $\theta _{2} $. Therefore the curve does not have an intersection with $\theta _{2}$-axis. The isocline in figure~\ref{fig_1b} is similarly defined for idea-II. The steady states of the two ideas are the intersections of the isoclines while plotting both of them in a single $\theta _{1}$-$\theta _{2}$ coordinate system \cite{ref-22,ref-30}, at which neither $\theta _{1} $ nor $\theta _{2} $ tends to increase or decrease. Note that the intersections  of the two isoclines (also known as equilibriums), together with the intersections between the isoclines and the $\theta _{1} $-$\theta _{2} $ axes, are the possible final states of ideas \cite{ref-22,ref-30}. A final state is \textit{stable} if other states close enough to it tend to move towards it; and \textit{unstable} otherwise.

\subsection{\label{subsec-3-2}Multiple coexistences of the competing ideas}
\begin{figure}
\begin{center}
\includegraphics[width=60mm,height=50mm]{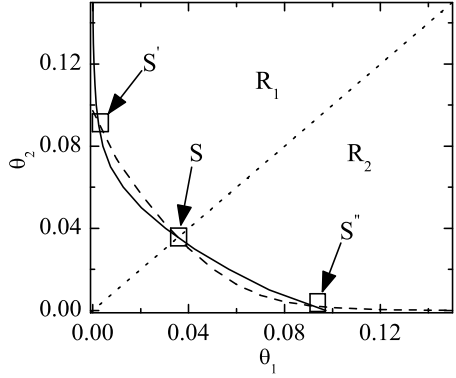} 
\end{center}
\caption{\label{fig_2}Diagram of the two ideas' zero-growth isoclines in the symmetrical influence model in an infinite scale-free network with exponent value $r=3$ and the minimum nodal degree $2$. The spreading rate and influential factor of both ideas are set to be $\lambda=0.3$ and $\gamma=0.15$ respectively. The solid and dashed curves represent the isoclines for idea-I and idea-II, respectively. The dotted line is the function $\theta_{1} =\theta_{2}$. Intersections of isoclines are highlighted by rectangles. There are two possible steady states $S^{'}$ and $S^{"}$ and an unstable state $S$.}
\end{figure}

We now prove the presence of multiple coexistent steady states when both ideas have non-exclusive influences. While accurate analysis on general scenarios of the model remains as a challenge, studies on some special cases may nevertheless reveal a few most important properties of the system. We consider a subclass of the general model where the two ideas have the same spreading rate (i.e., $\lambda _{1} =\lambda _{2} =\lambda$) and influential factor (i.e., $\alpha =\beta =\gamma$). We term such a case as the \textit{symmetrical influence} model. For this model, equations~(\ref{eq-3a})-(\ref{eq-3b}) can be re-written as 
\numparts 
\begin{eqnarray}
\theta _{1} =& \frac{1}{\langle k \rangle} \sum _{k}kP(k) \frac{  \lambda \left [ 1-(1-\theta _{1} )^{k} \right ] \left [ \gamma +(1-\gamma )(1-\theta _{2} )^{k} \right ] }{\displaystyle 1+\lambda \left [ 1-(1-\theta _{1} )^{k} \right ] \left [ \gamma +(1-\gamma )(1-\theta _{2} )^{k} \right ] } ,  \label{eq-6a} \\
\theta _{2} =& \frac{1}{\langle k \rangle} \sum _{k}kP(k) \frac{ \lambda \left [ 1-(1-\theta _{2} )^{k} \right ] \left [ \gamma +(1-\gamma )(1-\theta _{1} )^{k} \right ] }{\displaystyle 1+\lambda \left [ 1-(1-\theta _{2} )^{k} \right ] \left [ \gamma +(1-\gamma )(1-\theta _{1} )^{k} \right ] } , \label{eq-6b} 
\end{eqnarray} 
\endnumparts
which are coupled functions of $\theta_{1}$ and $\theta_{2}$. The isoclines of equations (\ref{eq-6a}) and (\ref{eq-6b}) are symmetric about the line $\theta _{1} =\theta _{2}$ as illustrated in figure \ref{fig_2}. Therefore, there are always an intersection point $S$ which can be denoted as $(\theta _{1} =\theta _{2} =\theta ^{*})$. From equations (\ref{eq-6a})-(\ref{eq-6b}), $\theta ^{*}$ can be expressed as the non-zero solution of
\begin{equation} \label{eq-7} 
\theta ^{*} =  \frac{1}{\langle k \rangle} \sum _{k}kP(k) \frac{ \lambda \left [ 1-(1-\theta ^{*} )^{k} \right ] \left [ \gamma +(1-\gamma )(1-\theta ^{*} )^{k} \right ] }{\displaystyle 1+\lambda \left [ 1-(1-\theta ^{*} )^{k} \right ] \left [ \gamma +(1-\gamma )(1-\theta ^{*} )^{k} \right ] } , 
\end{equation} 
which always exists for scale-free networks.

Taking derivative with respective to $\theta _{1} $ on both sides of equation~(\ref{eq-6a}), we have  the slope of isocline of idea-I expressed in terms of $\theta _{1} $ and $\theta _{2} $ as
\begin{equation} \label{eq-8} 
\fl \frac{d\theta _{2} }{d\theta _{1} } =g(\theta _{1} ,\theta _{2} )
= \frac{\displaystyle -\langle k \rangle+\sum _{k} k^{2} P(k) \frac{ \lambda (1-\theta _{1} )^{k-1} \left [\gamma +(1-\gamma )(1-\theta _{2} )^{k} \right ] }{\displaystyle \left ( 1+\lambda \left [ 1-(1-\theta _{1} )^{k} \right ] \left [ \gamma +(1-\gamma )(1-\theta _{2} )^{k} \right ] \right ) ^{2} }}{\displaystyle \sum _{k} k^{2} P(k) \frac{ \lambda \left [ 1-(1-\theta _{1} )^{k} \right ] (1-\gamma )(1-\theta _{2} )^{k-1} }{\displaystyle \left ( 1+\lambda \left [ 1-(1-\theta _{1} )^{k} \right ]  \left [ \gamma +(1-\gamma )(1-\theta _{2} )^{k} \right ] \right ) ^{2} }}.
\end{equation} 
Similarly, the slope of the isocline for equation~(\ref{eq-6b}) is
\begin{equation} \label{eq-9} 
\frac{d\theta _{2} }{d\theta _{1} } {\rm =}\frac{1}{g(\theta _{2} ,\theta _{1} )}.
\end{equation} 
The isocline of idea-I, as discussed earlier, does not intersect with $\theta _{2}$-axis since for any value of $\theta_{2}$, $\theta_{1}$ always has a non-trivial solution. Similarly, the isocline of idea-II does not intersect with $\theta_{1}$-axis, as illustrated in figure \ref{fig_2}. Therefore the \textit{sufficient} condition for the two isoclines to have at least two more intersections besides the intersection $S$ is that the slope of idea-I's isocline is greater than that of idea-II's isocline at the intersection point $S$, or in other words 
\begin{equation} \label{eq-10} 
g(\theta ^{*} ,\theta ^{*} )>\frac{1}{g(\theta ^{*} ,\theta ^{*} )}. 
\end{equation} 
From equation (\ref{eq-10}), we have 
\begin{equation} \label{eq-11} 
\sum _{k}k^{2} P(k) \frac{ \lambda (1-\theta ^{*} )^{k} }{\displaystyle \left ( 1+\lambda \left [ 1-(1-\theta ^{*} )^{k} \right ] \left [ \gamma +(1-\gamma )(1-\theta ^{*} )^{k} \right ] \right ) ^{2} }  < \langle k \rangle, 
\end{equation} 
where $\theta^{*}$ is defined in equation (\ref{eq-7}).

\begin{figure}
\begin{center}
\includegraphics[width=70mm,height=50mm]{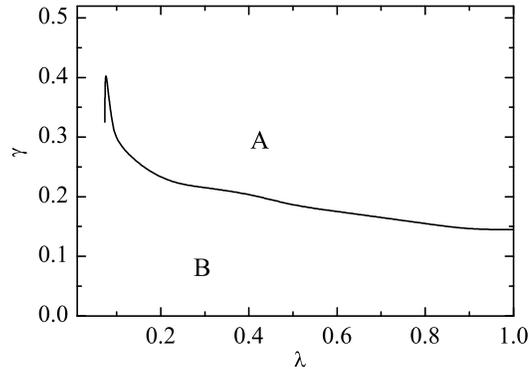} 
\end{center}
\caption{\label{fig_3}Schematic phase transition diagram of ideas in terms of spreading rate and influential factor in an infinite scale-free network with exponent value $r=3$ and the minimum nodal degree $2$. In the region $B$, it can be proved that the two ideas have multiple coexistences.}
\end{figure}

The region of $\lambda $ and $\gamma $ where the sufficient condition is satisfied is illustrated as region $B$ in figure~\ref{fig_3}. In this region, there are at least two stable steady states that the ideas can finally reach, as illustrated in figure~\ref{fig_2}: though the two ideas always coexist, their final densities are not unique, but controlled by their initial densities instead. Specifically, the idea with a relatively higher initial density reaches a higher final density as well. This can also be observed in figure~\ref{fig_2}. For example, assume that the initial densities of the ideas lie in region $R_{1}$ (above the dotted line), in which idea-II has a relatively higher density, the system will converge to the steady state $S^{'} $ in region $R_{1} $, where eventually idea-II still has a relatively higher density. In the rare case where the two ideas have exactly the same initial densities, they reach an \textit{unstable} state with the same density at the end, e.g., the $S$ state in figure~\ref{fig_2}. Note that as shown in figure~\ref{fig_3}, when $\lambda $ decreases, the critical value of $\gamma $ for satisfying the sufficient condition increases, meaning that weaker transmissibility tends to make multiple coexistences easier to happen. When the spreading rate $\lambda $ approaches zero, however, there is a sharp drop in the critical value of $\gamma $. This can be explained: at low transmissibility, the densities of both ideas are very low. The neighborhood influence has to be very strong in order to have nontrivial effects on the spreading of the competitors leading to multiple coexistences.  Such a result can also be derived from equations~(\ref{eq-7}) and (\ref{eq-11}). From equation~(\ref{eq-7}), in scale-free networks, when $\lambda \to 0$, $\theta ^{*} \to 0$ \cite{ref-20}. Therefore, regardless the value of influential factor $\gamma $, the left side of equation~(\ref{eq-11}) can be approximated as $\lambda \sum _{k}k^{2} P(k) $, which cannot be guaranteed to be less than $\langle k \rangle$.

Note that the presence of multiple coexistences in region $B$ is proved by using the sufficient condition, which does not theoretically eliminate the possibility that multiple coexistences may also occur in certain areas in region $A$. In our simulation, however, coexistence in region $A$ has never been observed.

\subsection{\label{subsec-3-3} Numerical simulations and discussions}

\begin{figure}
\begin{center}
\subfigure[]{\label{fig_4a}
\includegraphics[width=70mm,height=50mm]{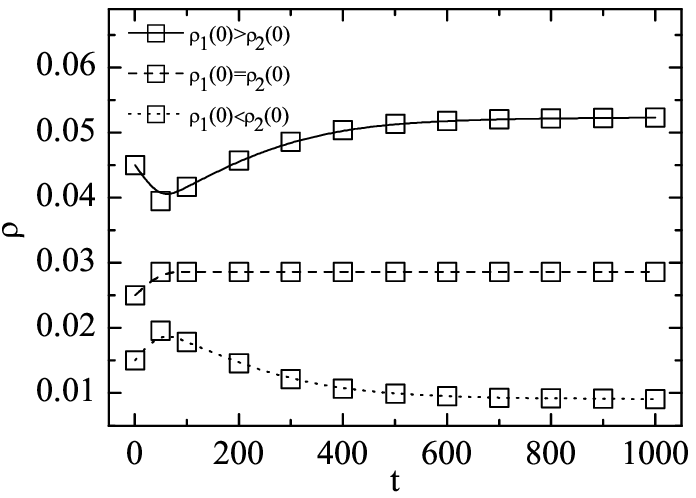}}
\subfigure[]{\label{fig_4b}
\includegraphics[width=70mm,height=50mm]{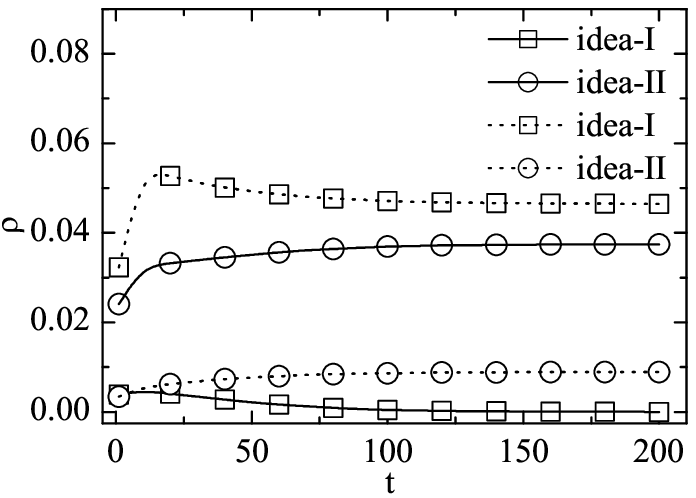}}
\end{center}
\caption{\label{fig_4}Time evolution of the densities of competing ideas. The network is of a finite size of $10,000$ nodes, an exponent value of $r=3$ and the minimum nodal degree $2$. In (a), the spreading rate and the influential factor of both ideas are set to be $\lambda =0.3$ and $\gamma =0.25$, respectively. The initial density of idea-II is always set as $0.025$. The solid, dashed and dotted lines represent the time-evolution densities of idea-I with initial density higher than, equal to and lower than that of idea-II, respectively.  In (b), the spreading rates of idea-I and idea-II are $\lambda _{1} =0.3$ and $\lambda _{2} =0.25$, and their influential factors are $\alpha =0.05$ and $\beta =0.45$ respectively. The solid and dotted lines represent two sets of simulation results with different initial densities. Presented results are averaged over $100$ realizations.}
\end{figure}

Numerical results demonstrating the presence of multiple coexistences in symmetrical and asymmetrical influential agent models are presented in figure~\ref{fig_4}. Specifically, the network is generated by adopting the uncorrelated configuration model (UCM) in \cite{ref-31}, of a finite size of $10,000$ nodes, an exponent value of $3$ and the minimum nodal degree of $2$. Initially, a certain number of randomly selected nodes are infected by idea-I and idea-II, respectively. Then in each discrete time step, a node exposed to idea-I can be  infected by the idea at a rate of $\nu _{1} =\lambda _{1} \delta _{1} $ if and only if there exist only idea-I infected nodes but no idea-II infected nodes among its adjacent nodes. Corresponding assumption applies to idea-II. When there exist both idea-I and idea-II infected nodes in the neighborhood, a node exposed to both ideas can be infected by the two ideas at rates of $\nu _{1} =\alpha \lambda _{1} \delta _{1} $  and $\nu _{2} =\beta \lambda _{2} \delta _{2} $, respectively. Without loss of generality, we typically adopted $\delta _{1} =\delta _{2} =0.5$ \cite{ref-17, ref-20} in our simulations. Repeat the above procedure until reaching a steady state. The simulation results are averaged over at least $100$ realizations.

Figure~\ref{fig_4a} shows the time dynamics of the density of idea-I in the symmetrical influence model, where the initial density of idea-I is higher than, equal to, and lower than that of idea-II, respectively. We see that idea-I can have very different final densities depending on its relative initial densities against that of idea-II, which matches our analytical results. Figure~\ref{fig_4b} illustrates the presence of multiple coexistences in more general cases where the two ideas have different spreading rates and influential factors. Overall, the initial densities of the ideas appear to play an important role in determining the steady states: a relative higher density at the initial stage helps the idea suppress its competitor to a low level and gain an advantageous position at the steady state, and vice versa.

\begin{figure}
\begin{center}
\includegraphics[width=70mm,height=50mm]{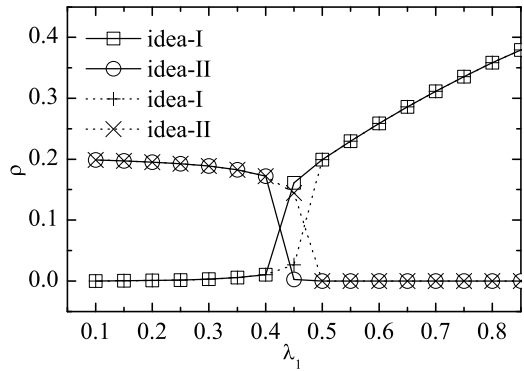} 
\end{center}
\caption{\label{fig_5}Densities of competing ideas at stationary state. The simulation results are on top of a $10,000$-node scale-free network with exponent $r=3$ and the minimum nodal degree $2$. The product $\lambda _{1} \alpha $ is fixed to be $0.1$. For idea-II, its spreading rate $\lambda _{2} $ and influential factor $\beta $ are fixed to be $0.5$ and $0.1$ respectively. The solid and dotted lines represent two different sets of simulations with initial densities of ideas $\rho _{1} (0)=0.5$ and $\rho _{2} (0)=0.01$ (solid lines) and $\rho _{1} (0)=0.01$ and $\rho _{2} (0)=0.5$(dotted lines), respectively. Simulated data is average over at least 100 realizations.}
\end{figure}

It is of interest to figure out, when subject to limited resources, whether it is more effective to increase the spreading rate or to weaken the neighborhood influence imposed by the competitor idea. Specifically, we consider the case where the product of an idea's spreading rate and influence factor is a constant. By re-writing the term $\lambda _{1} [ \alpha +(1-\alpha )(1-\theta _{2} )^{k} ]$ as $\lambda _{1} \alpha [1-(1-\theta _{2} )^{k} ]+\lambda _{1} (1-\theta _{2} )^{k} $ in equation~(\ref{eq-3a}), we have that
\begin{equation} \label{eq-12} 
 \fl \theta _{1}  =\frac{1}{\langle k \rangle} \sum _{k} kP(k) \frac{(1-\theta _{1} )^{k} \left ( \lambda _{1} \alpha \left [ 1-(1-\theta _{2} )^{k} \right ] +\lambda _{1} (1-\theta _{2} )^{k} \right ) }{\displaystyle 1+(1-\theta _{1} )^{k} \left ( \lambda _{1} \alpha \left [ 1-(1-\theta _{2} )^{k} \right ] +\lambda _{1} (1-\theta _{2} )^{k}  \right ) } 
\end{equation} 
which implies that when $\lambda _{1} \alpha $ is of a constant value and all the other parameters remain fixed, a higher spreading rate $\lambda _{1} $ results in a larger population accepting idea-I at the steady state. Figure~\ref{fig_5} presents some supportive numerical simulation results. The product $\lambda _{1} \alpha $ is fixed at $0.1$, and the spreading rate and influential factor of idea-II are fixed at $0.5$ and $0.1$ respectively. To still illustrate multiple coexistences, two different sets of numerical simulations with initial densities $\rho _{1} (0)=0.5$ and $\rho _{2} (0)=0.01$ and $\rho _{1} (0)=0.01$ and $\rho _{2} (0)=0.5$  (reflecting the cases where $\rho _{1} (0)\gg \rho _{2} (0)$ and $\rho _{1} (0)\ll \rho _{2} (0)$ respectively) have been conducted. As we can see, the final density of population accepting idea-I at the steady state increases with $\lambda_{1}$. At the beginning, the increasing speed is slow; and then it becomes much faster when $\lambda_{1}$ is high enough. Multiple coexistences of the two ideas can also be observed in figure~\ref{fig_5}, e.g., when $\lambda _{1} =0.45$.  The simulation results confirm that it is more effective to increase spreading rate rather than weakening the neighborhood influence of the competitor idea. Note that due to the finite-size effect in numerical simulation, idea-II is driven to be virtually extinct when the spreading rate of idea-I is very high.

\section{\label{sec-4} Exclusive influence vs exclusive influence}
\subsection{\label{subsec-4-1} Non-coexistence of competing ideas with exclusive influences}

In this section, we prove that two competing ideas with exclusive influences can never stably coexist in any scale-free networks. Specifically, we show that the necessary condition for two competing ideas to coexist cannot be satisfied when both of the ideas are with exclusive influences. 

Still denote the spreading rates of the two ideas as $\lambda _{1} $ and $\lambda _{2}$, respectively. By setting $\alpha =\beta =0$ in equations (\ref{eq-3a})-(\ref{eq-3b}), we have that $\theta _{1}$ and $\theta _{2}$ satisfy
\numparts 
\begin{eqnarray}
\theta _{1} = \frac{1}{k} \sum _{k}kP(k)\frac{\lambda _{1} \left [ 1-(1-\theta _{1} )^{k} \right ] (1-\theta _{2} )^{k} }{\displaystyle 1+\lambda _{1} \left [ 1-(1-\theta _{1} )^{k} \right ] (1-\theta _{2} )^{k} },  \label{eq-13a} \\
\theta _{2} = \frac{1}{k}  \sum _{k}kP(k)\frac{\lambda _{2} \left [ 1-(1-\theta _{2} )^{k} \right ] (1-\theta _{1} )^{k} }{\displaystyle 1+\lambda _{2} \left [ 1-(1-\theta _{2} )^{k} \right ] (1-\theta _{1} )^{k} } . \label{eq-13b}
\end{eqnarray}
\endnumparts 

For the two competing ideas to stably coexist, neither of them should be driven out by the other. Therefore, if we denote the final state of idea-I with spreading rate $\lambda _{1} $ when there is only idea-I in the system as $\theta _{1}^{A} $ and  the minimum density of idea-I to exclude idea-II from the system as $\theta _{1}^{A^{'} } $, to avoid having idea-II being driven out, we shall have $\theta _{1}^{A} < \theta _{1}^{A^{'} } $. Similarly, by denoting the minimum density of idea-II to exclude idea-I from the system as $\theta _{2}^{B} $ and the final state of idea-II with spreading rate $\lambda _{2} $ when there is only idea-II in the system as $\theta _{2}^{B^{'} } $, we shall have $\theta _{2}^{B} > \theta _{2}^{B^{'} } $ to keep idea-I from being driven out. Thereby, $\theta _{2}^{B} >\theta _{2}^{B^{'} } $ and $\theta _{1}^{A} <\theta _{1}^{A^{'} } $ are necessary conditions for coexistence. Below we prove that such conditions are not satisfied. 

Similar to that for  equations~(\ref{eq-3a}) and (\ref{eq-3b}), equations~(\ref{eq-13a}) and (\ref{eq-13a}) can be plotted into two separate isoclines as functions of $\theta _{1} $ and $\theta _{2} $. Different from that in figure~\ref{fig_1}, however, each isocline intersects with both the $\theta _{1} $- and $\theta _{2} $-axes. Specifically, assume that for idea-I the intersection points are $A$ and $B$. We can have that their coordinates are $(\theta _{1}^{A}, 0)$ and $(0, \theta _{2}^{B} )$ respectively, where $\theta _{1}^{A} $ and $\theta _{2}^{B} $ are defined as above. Similarly, denote the intersecting points of the isocline for idea-II with  $\theta _{1} $- and $\theta _{2} $- axes as $A^{'}$ and $B^{'} $. We have that their coordinates shall be $(\theta _{1}^{A^{'} }, 0)$ and $(0, \theta _{2}^{B^{'} } )$, respectively. 

\begin{figure}
\begin{center}
\subfigure[]{\label{fig_6a}
\includegraphics[width=50mm,height=50mm]{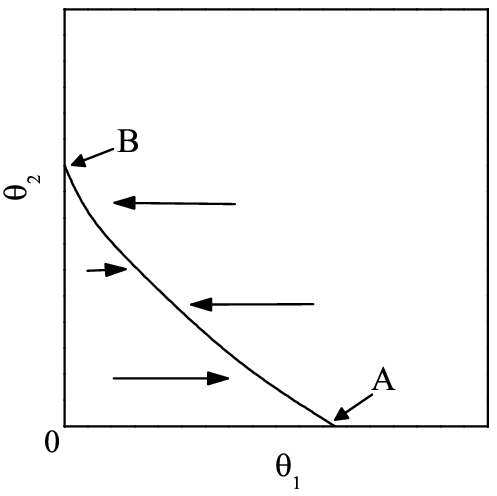}}
\subfigure[]{\label{fig_6b}
\includegraphics[width=50mm,height=50mm]{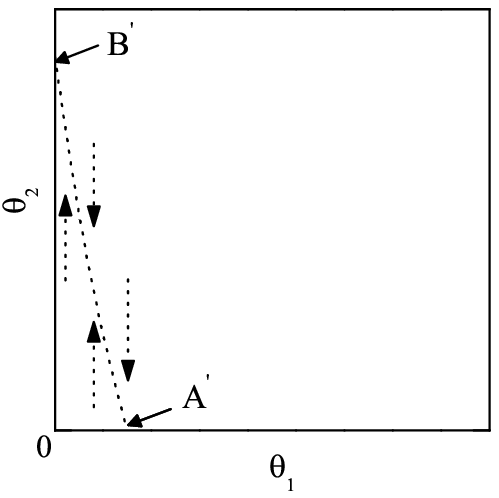}}
\end{center}
\caption{\label{fig_6}Schematic phase plane diagrams of the two ideas' zero-growth isoclines in an infinite scale-free network with exponent $r=3$ and the minimum degree $2$. Both of the two ideas have exclusive influences. The spreading rates are set to be $\lambda _{1} =0.25$ and $\lambda _{2} =0.3$, respectively. Arrows indicate the moving directions of $\theta _{1}$ ($\theta _{2} $) when $\theta _{2} $ ($\theta _{1} $) holds as a constant.}
\end{figure}

From equations~(\ref{eq-13a}) and (\ref{eq-5}) we have that $\theta _{1}^{A} $ and $\theta _{2}^{B} $ are solutions of the equations
\numparts 
\begin{eqnarray}
\theta _{1} =\frac{1}{\langle k \rangle}  \sum _{k}kP(k) \frac{\lambda _{1} \left [ 1-(1-\theta _{1} )^{k} \right ] }{\displaystyle 1+\lambda _{1} \left [ 1-(1-\theta _{1} )^{k} \right ] } 
\label{eq-14a} \\
\fl \mbox{and} \nonumber \\
1=\frac{\lambda _{1}}{\langle k \rangle} \sum _{k}k^{2} P(k)(1-\theta _{2} )^{k} , \label{eq-14b}
\end{eqnarray}
\endnumparts 
respectively. Specifically, equation~(\ref{eq-14a}) comes from equation~(\ref{eq-13a}) by letting $\theta _{2} =0$. It describes the stationary state of a single epidemic with a spreading rate $\lambda _{1} $ in a random network with the nodal-degree distribution $P(k)$. In scale-free networks with exponent $r\le 3$, $\theta _{1}^{A} $ yielded from equation~(\ref{eq-14a}) is non-zero and unique for any given non-zero spreading rate $\lambda _{1}$ \cite{ref-20,ref-29}. Equation~(\ref{eq-14b}) comes from  equation~(\ref{eq-5}) showing the critical condition of the inequality. Solving equation~(\ref{eq-14b}), we get the value of $\theta _{2}^{B} $, which is the minimum value of $\theta _{2} $ for idea-II to drive out idea-I from the system \cite{ref-22,ref-30}. Equation~(\ref{eq-14b}) can be re-written as $\lambda _{1} = \langle k \rangle / (\sum _{k}k^{2} P(k)(1-\theta _{2} )^{k} )$. Since  $ \langle k \rangle / \sum _{k}k^{2} P(k) =0$ in an infinite scale-free network and $\lambda _{1} >0$, equation~(\ref{eq-14b}) always has a positive solution $0<\theta _{2}^{B} <1$. Further noting that $\sum _{k}k^{2} P(k)(1-\theta _{2} )^{k}  $ is a monotonically decreasing function of $\theta _{2} $ in $[0, 1]$, we have that equation~(\ref{eq-14b}) has a unique positive solution $\theta _{2}^{B} $. The isocline of idea-I is plotted in figure~\ref{fig_6a}. Below we compare the values of $\theta _{1}^{A} $ and $\theta _{2}^{B} $. 

It can be observed from equations (\ref{eq-14a}) and (\ref{eq-14b}) that both $\theta_{1}$ and $\theta _{2}$ increase with $\lambda_{1}$. Let $\theta_{1}$ in equation (\ref{eq-14a}) and $\theta _{2}$ in equation (\ref{eq-14b}) be equal to each other (denoted as $\theta$) and solve the corresponding values of $\lambda_{1}$ (denoted as $\lambda _{a}$ and $\lambda _{b}$ respectively). We have 
\begin{equation} \label{eq-15} 
\lambda _{a}^{-1} -\lambda _{b}^{-1} = \frac{1}{\langle k \rangle} \sum _{k}kP(k) \frac{\displaystyle  1-(1+k\theta +k\lambda _{a} \theta )\ell +k\lambda _{a} \theta \ell ^{2} }{\displaystyle  \theta \big [ 1+\lambda _{a} (1-\ell ) \big ] }  ,                  
\end{equation} 
where $\ell =(1-\theta )^{k}$. Consider the term $1-(1+k\theta +k\lambda _{a} \theta )\ell +k\lambda _{a} \theta \ell ^{2} =f(\ell )$ as a parabola function of $\ell$. Since the axis of symmetry of the parabola is at $e=\frac{1}{2} +\frac{1}{2\lambda _{a} } +\frac{1}{2k\lambda _{a} \theta } >1$, for $\ell \in \left[0, 1\right]$, $f(\ell )$ reaches the minimum value at $\ell =1$ or equivalently $\theta =0$. Therefore, for any value of $\theta $, we have $\lambda _{a}^{-1} -\lambda _{b}^{-1} >0$. In other words, $\theta _{2}^{B} $ is always smaller than $\theta _{1}^{A}$ for any given $\lambda _{1}$.

Now consider $\theta _{1}^{A^{'} } $ and $\theta _{2}^{B^{'} } $. From equations~(\ref{eq-5}) and (\ref{eq-13b}), $\theta _{1}^{A^{'} } $ and $\theta _{2}^{B^{'} } $ satisfy 
\numparts
\begin{eqnarray}
1=\frac{\lambda _{2}}{\langle k \rangle} \sum _{k}k^{2} P(k)(1-\theta _{1} )^{k} \label{eq-16a}\\
 \fl \mbox{and} \nonumber \\ 
\theta _{2} =\frac{1}{\langle k \rangle } \sum _{k} kP(k) \frac{\lambda _{2} \left [ 1-(1-\theta _{2} )^{k} \right ] }{\displaystyle 1+\lambda _{2} \left [ 1-(1-\theta _{2} )^{k} \right ] }, \label{eq-16b}
\end{eqnarray}
\endnumparts
respectively. Similar to that from our discussions above, we have that $\theta _{1}^{A^{'} } < \theta _{2}^{B^{'} } $. The isocline of idea-II is plotted in figure~\ref{fig_6b}.

The conclusions $\theta _{2}^{B} <\theta _{1}^{A} $ and $\theta _{1}^{A^{'} } <\theta _{2}^{B^{'} } $ contradicts the necessary condition of coexistence that $\theta _{2}^{B} >\theta _{2}^{B^{'} } $ and $\theta _{1}^{A} <\theta _{1}^{A^{'} } $. Therefore, the two ideas cannot stably coexist.

\subsection{\label{subsec-4-2} Founder control, exclusion and phase transition in between}

With the understanding that stable coexistence of the two ideas is impossible, we proceed to analyze the possible final states. 

When the two ideas are with comparable spreading rates, or more specifically when $\theta _{1}^{A} >\theta _{1}^{A^{'} } $ and $\theta _{2}^{B} <\theta _{2}^{B^{'} } $, either of them can drive out the other one. The final states of them are determined by their initial densities, i.e., they are in the state of the founder control. Consider the first inequality $\theta _{1}^{A} >\theta _{1}^{A^{'} } $. Since  $\theta _{1}^{A} $ and $\theta _{1}^{A^{'} } $ are solutions of equations~(\ref{eq-14a}) and (\ref{eq-16a}) respectively, by replacing $\theta _{1}^{A^{'} } $ with $\theta _{1}^{A} $ in equation~(\ref{eq-16a}), we have that
\begin{equation} \label{eq-17} 
\lambda _{2} < \lambda _{2,c} = \frac{\langle k \rangle }{\sum _{k}k^{2} P(k)(1-\theta _{1}^{A} ) ^{k} },              
\end{equation} 
where $\lambda _{2,c} $ is the boundary spreading rate of idea-II above which the founder control cannot happen. From $\theta _{2}^{B} <\theta _{2}^{B^{'} } $, similarly, we have
\begin{equation} \label{eq-18} 
\lambda _{1}  <\lambda _{1,c} =\frac{\langle k \rangle}{\sum _{k}k^{2} P(k)(1-\theta _{2}^{B^{'} } ) ^{k}},                           
\end{equation} 
where $\lambda _{1,c} $ is the boundary spreading rate of idea-I. To have founder control, both equations~(\ref{eq-17}) and (\ref{eq-18}) need to be satisfied.

\begin{figure}
\begin{center}
\subfigure[]{\label{fig_7a}
\includegraphics[width=60mm,height=50mm]{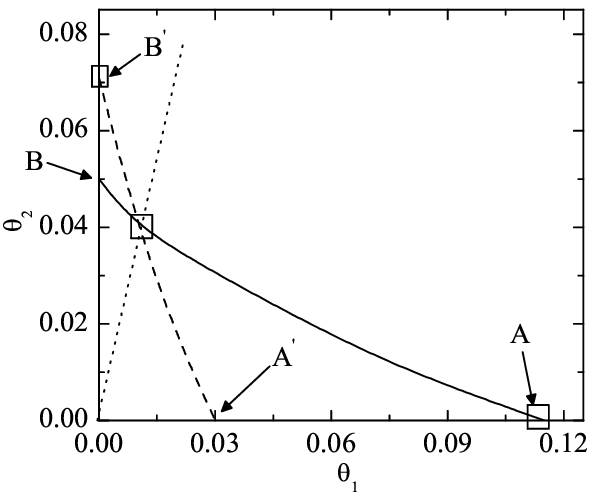}}
\subfigure[]{\label{fig_7b}
\includegraphics[width=90mm,height=50mm]{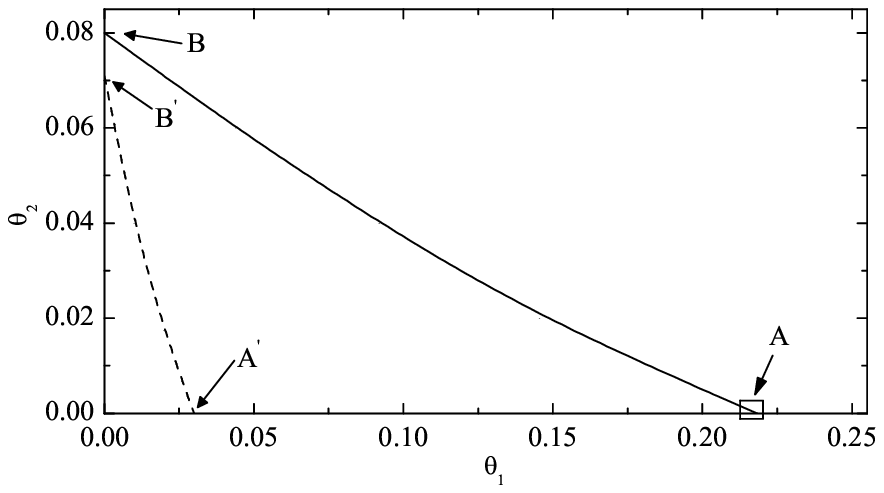}}
\end{center}
\caption{\label{fig_7}Diagram of the final states of the two competing ideas, both with exclusive influences, in an infinite scale-free network with exponent $r=3$ and the minimum degree 2, described by isoclines as functions of $\theta _{1} $ and $\theta _{2} $. The solid and dashed curves represent the isoclines for idea-I and idea-II, respectively. In (a), the spreading rates of idea-I and idea-II are $\lambda _{1} =0.3$ and $\lambda _{2} =0.25$ respectively, and the competing ideas are under founder control. The dotted line in (a) represents the case where ideas can unstably coexist. In (b), the spreading rates of idea-I and idea-II are $\lambda _{1} =0.4$ and $\lambda _{2} =0.25$, respectively. The ideas are in the state of exclusion.}
\end{figure}

Figure \ref{fig_7a} shows an example case of founder control in an infinite scale-free network with exponent $r=3$ and the minimum nodal degree $2$ where the spreading rates of idea-I and idea-II are $0.3$ and $0.25$ respectively. We observe that if the initial values of $\theta _{1}$ and $\theta _{2}$ are located above the dotted line, idea-I will be driven out while idea-II will persist and reach its steady state at point $B^{'}$. If the initial values lie below the dotted line, however, idea-II will be driven out while idea-I will persist and reach its steady state at point $A$. If their initial values happen to be on the dotted line, the two ideas will unstably coexist, which is not sustainable since even small fluctuations can easily destroy the coexistence. 

When the two ideas are of rather different spreading rates, one idea may drive out the other regardless of their initial densities, i.e., the ideas are in the state of \textit{exclusion}. Without loss of generality, we consider the case where idea-I has a higher spreading rate. To exclude idea-II, it requires that $\theta _{1}^{A} >\theta _{1}^{A^{'} } $ and $\theta _{2}^{B} >\theta _{2}^{B^{'} } $. Since $\theta _{1}^{A} >\theta _{2}^{B} $ and $\theta _{2}^{B^{'} } >\theta _{1}^{A^{'} } $ are always valid, only $\theta _{2}^{B} >\theta _{2}^{B^{'} } $ needs to be satisfied where $\theta _{2}^{B} $ is the solution of equation~(\ref{eq-13b}) and $\theta _{2}^{B^{'} } $is the solution of equation~(\ref{eq-16b}). Bringing the condition $\theta _{2}^{B} >\theta _{2}^{B^{'} } $ into equation~(\ref{eq-14b}), we have that to exclude idea-II, the spreading rate of idea-I needs to satisfy
\begin{equation} \label{eq-19} 
\lambda _{1} >\lambda _{1,c} = \frac{\langle k \rangle} {\sum _{k}k^{2} P(k)(1-\theta _{2}^{B^{'} } ) ^{k}} ,                              
\end{equation} 
where $\theta _{2}^{B^{'} }$ is the solution of equation~(\ref{eq-16b}) with the given spreading rates $\lambda _{2} $. Figure~\ref{fig_7b} shows the isoclines of ideas for an example case where spreading rates of idea-I and idea-II are 0.4 and 0.25, respectively. In the steady state, idea-II is always driven out by idea-I. 

\begin{figure}
\begin{center}
\includegraphics[width=50mm,height=50mm]{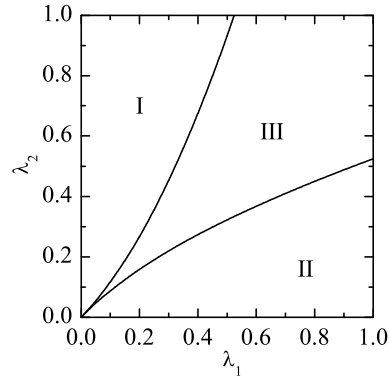} 
\end{center}
\caption{\label{fig_8}Schematic phase diagram, in terms of spreading rates, of two competing ideas with exclusive influences in an infinite scale-free network with exponent value $r=3$ and the minimum nodal degree $2$. In region I, idea-II persists while idea-I dies out from the network; in region II, idea-I persists while idea-II dies out from the network; region I and II represent the areas where exclusion happens. Region III represents the area of founder control.}
\end{figure}

The above analysis shows that there exist different regions of spreading rates leading to founder control or exclusion. An example of different spreading rates leading to different steady states is illustrated in figure~\ref{fig_8}. The boundary between region II and region III comes from equations~(\ref{eq-18}) and (\ref{eq-19}). It is formed by the critical values of $\lambda _{1,c} $ corresponding to different values of $\lambda _{2} $. The boundary between regions I and III is formed by the critical values of $\lambda _{2,c} $ corresponding to different values of $\lambda _{1} $. Region III represents the spreading rates leading to founder control, while regions I and II represent the spreading rates leading to exclusion. The solid curves represent the critical phase transition between these regions. Figure~\ref{fig_8} also verifies that stable coexistence does not happen  between two competing ideas both with exclusive influence.

\subsection{ \label{subsec-4-3}Numerical simulations}

\begin{figure}
\begin{center}
\subfigure[]{\label{fig_9a}
\includegraphics[width=70mm,height=50mm]{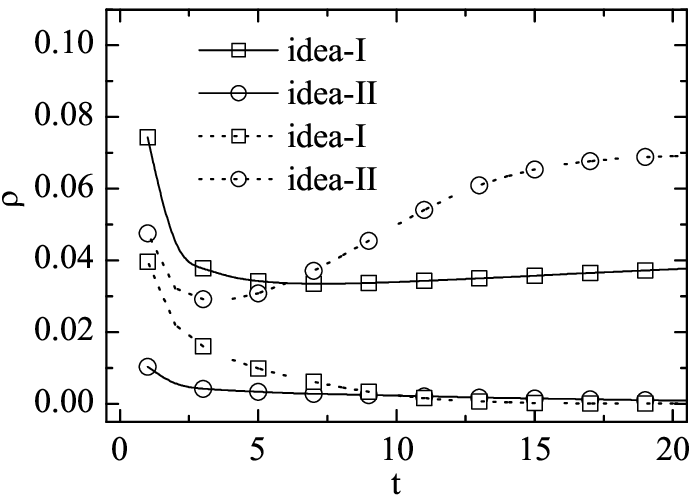}}
\subfigure[]{\label{fig_9b}
\includegraphics[width=70mm,height=50mm]{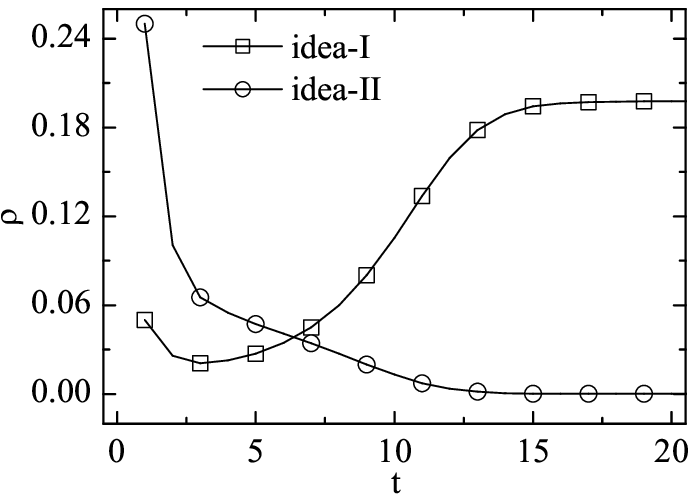}}
\end{center}
\caption{\label{fig_9}Time evolution of the densities of two competing ideas with exclusive influences. The network has a finite size of $10,000$ nodes, an exponent value of $r=3$ and the minimum nodal degree $2$. Square boxes represent the densities of idea-I while circles represent those of idea-II. Solid and dotted curves in (a) represent two sets of simulations with different initial densities of the competing ideas. In (a), the spreading rates of idea-I and idea-II are $\lambda _{1} =0.25$ and $\lambda _{2} =0.3$, respectively; in (b), they are $\lambda _{1} =0.5$ and $\lambda _{2} =0.3$, respectively. Data shown is averaged over at least 100 realizations.}
\end{figure}

A few examples of the time evolution of two competing ideas both with exclusive influences are illustrated in figure~\ref{fig_9}. Simulations are conducted on the same network model as adopted in Section \ref{sec-3}. Figure~\ref{fig_9a} demonstrates example cases of founder control where the steady states of the two ideas depend on their initial densities. In contrast, idea-I in figure~\ref{fig_9b} always drives out idea-II even when the initial density of idea-II is very high, which demonstrates the exclusion case when the difference between the two ideas' spreading rates is large enough. Such simulation results support our analysis. 

\section{\label{sec-5} Exclusive influence vs non-exclusive influence}

We have discussed the cases where both of the competing ideas have the same type of neighborhood influences. Now we study the dynamics and phase transition of two competing ideas with exclusive and non-exclusive influences respectively.

\subsection{\label{subsec-5-1}Exclusion, multiple endemic states, coexistence and phase-transition}

Without loss of generality, we assume that idea-I has exclusive influence, while idea-II imposes only non-exclusive influence on idea-I. The influential factors are therefore $\alpha >0$ for idea-I and $\beta =0$ for idea-II. By letting $\alpha >0$ and $\beta =0$ in equations~(\ref{eq-1a})-(\ref{eq-3b}), we have that the endemic states of  ideas $\theta _{1} $ and $\theta _{2} $ have to fulfill
\numparts
\begin{eqnarray}
\theta _{1} =\frac{1}{\langle k \rangle } \sum _{k} kP(k) \frac{\lambda _{1} \left [ 1-(1-\theta _{1} )^{k} \right ] \left [ \alpha +(1-\alpha )(1-\theta _{2} )^{k} \right ] }{\displaystyle  1+\lambda _{1} \left [ 1-(1-\theta _{1} )^{k} \right ] \left [ \alpha +(1-\alpha )(1-\theta _{2} )^{k} \right ] }  , \label{eq-20a}\\
\theta _{2} =\frac{1}{\langle k \rangle }\sum _{k}kP(k) \frac{ \lambda _{2} \left [ 1-(1-\theta _{2} )^{k} \right ] (1-\theta _{1} )^{k} }{\displaystyle  1+\lambda _{2} \left [ 1-(1-\theta _{2} )^{k} (1-\theta _{1} )^{k} \right ] }  . \label{eq-20b}
\end{eqnarray}
\endnumparts
From the analysis in Sections~\ref{sec-3} and \ref{sec-4}, the schematic isocline for the function of equation (\ref{eq-20a}) has the same properties as the one shown in figure \ref{fig_1a}, e.g., it has no intersection with the $\theta _{2}$-axis. Similarly, equation (\ref{eq-20b}) defines an isocline with the same properties as those of the isocline in equation (\ref{eq-13b}), which does have an intersection with $\theta_{1}$-axis. By analyzing isoclines, we show that the possible stationary states of the two competing ideas include exclusion where one idea definitely drives out the other, multiple endemic states where there are multiple stationary states including exclusion and coexistence depending on the initial densities of the ideas, and stable coexistence. We study the conditions for obtaining each of these stationary states and the phase-transition criteria between them. 

\begin{figure}
\begin{center}
\subfigure[]{\label{fig_10a}
\includegraphics[width=50mm,height=50mm]{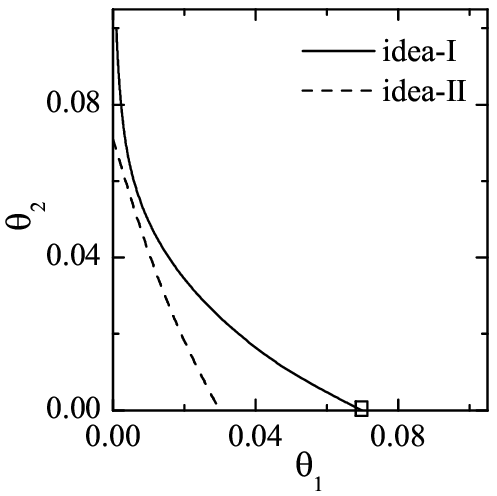}}
\subfigure[]{\label{fig_10b}
\includegraphics[width=50mm,height=50mm]{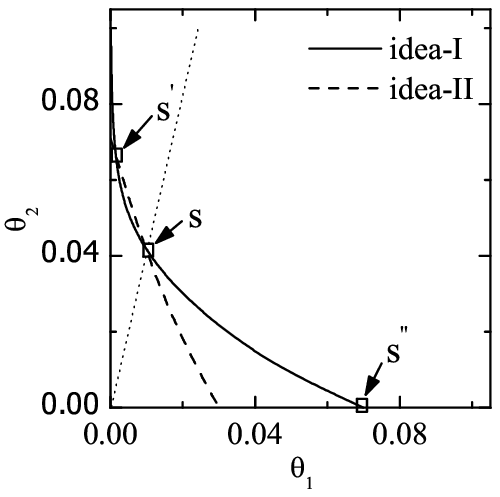}} \\
\subfigure[]{\label{fig_10c}
\includegraphics[width=50mm,height=50mm]{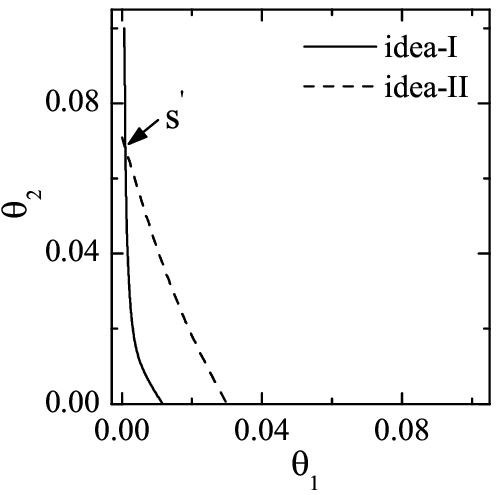}}
\end{center}
\caption{\label{fig_10} Diagram of the zero-growth isoclines for competing ideas with exclusive and non-exclusive influences respectively in an infinite scale-free network with exponent value 3 and the minimum nodal degree $2$. The spreading rate of idea-II is always set to be $\lambda_{2}=0.25$. The spreading rate of idea-I is $\lambda_{1}=0.26$ in (a) and (b), while the influential factors are $\alpha=0.2$ in (a) and $\alpha=0.1$ in (b) respectively. In (c), the spreading rate and influential factor of idea-I are $\lambda_{1}=0.15$ and $\alpha=0.5$ respectively.}
\end{figure}

As we discussed in Section~\ref{sec-4}, where both of the ideas impose exclusive influences (i.e., $\alpha=\beta=0$) to each other, Idea-II is always excluded when the spreading rate of idea-I satisfies $\lambda _{1} >\lambda _{1,c} $ as defined in equation~(\ref{eq-19}). Conversely, from equations~(\ref{eq-14a}) and (\ref{eq-16b}), we can have the critical spreading rate $\lambda _{1,c}^{'} $ of idea-I, below which idea-I is always driven out. These critical rates are also of importance in determining the steady state of competing ideas with exclusive and nonexclusive influences respectively. Since it is obvious that for $\lambda _{1} \ge \lambda _{1,c} $, idea-II is always driven out, below we discuss the other two different cases where $\lambda _{1,c}^{'} \le \lambda _{1} <\lambda _{1,c} $ and $\lambda _{1} <\lambda _{1,c}^{'} $, respectively.

For $\lambda _{1,c}^{'} \le \lambda _{1} <\lambda _{1,c} $, when the influential factor $\alpha$ is small enough, e.g., considering the extreme case where it approaches zero, the two competing ideas can have multiple endemic states. When $\alpha $ is high enough, on the other hand, idea-I drives out idea-II. Such two different cases correspond to that the system of equations~(\ref{eq-20a}) and (\ref{eq-20b}) has multiple or no non-zero solutions of $\left(\theta _{1}, ~\theta _{2} \right)$ respectively. Denote \textit{the critical influential factor} $\alpha_{c} $ as the value of $\alpha $ where the system of equations (\ref{eq-20a}) and (\ref{eq-20b}) has exactly one non-zero solution; $\alpha_{c} $ can be numerically calculated by employing the \textit{dual simplex method} \cite{ref-27}. Below we show that for $\lambda _{1,c}^{'} \le \lambda _{1} <\lambda _{1,c} $ and $\alpha<\alpha_{c} $, there are multiple endemic states, among which at least one is coexistence and another is idea-I excludes idea-II.

To simplify the discussions, we illustrate an example case where the spreading rates of idea-I and idea-II are set to be $0.26$ and $0.25$ respectively. The underlying network is an infinite scale-free network with exponent $r=3$ and the minimum nodal degree $2$. For such a case, numerical calculations show that $\lambda _{1,c} \simeq 0.354$, $\lambda _{1,c}^{'} \simeq 0.191$ and $\alpha_{c} \simeq 0.151$. The isoclines of the ideas are illustrated in figure~\ref{fig_10}. If the influential factor $\alpha$ is high enough, e.g. $\alpha =0.2$ as that in figure~\ref{fig_10a}, idea-II is always driven out. If the influential factor is low, e.g., $\alpha =0.1$ as that in figure~\ref{fig_10b}, there are multiple possible stationary states. Now we show evidence of the existence of multiple stationary states. Denote $(\theta _{1}^{*} ,~\theta _{2}^{*} )$ as the unique intersection of the two isoclines when $\alpha=\alpha _{c} $. Since the isocline of idea-I approaches the $\theta _{2} $-axis yet never intersects with it, we have that the two isoclines are tangential to each other (otherwise another intersection point must exist). Therefore, for $\alpha < \alpha _{c}$, there exist at least two intersections $S$ and $S^{'}$ as illustrated in figure~\ref{fig_10b}. At least one of them corresponds to a steady state. Together with another steady state $S^{"} $ where idea-I excludes idea-II, there are at least two different steady states. The existence of multiple endemic steady states is therefore proved. 

For $\lambda _{1} <\lambda _{1,c}^{'} $, the analysis is relatively much simpler. We can easily have that idea-I will never be driven out from the system since $\alpha >0$.  On the other hand, since $\lambda _{1} <\lambda _{1,c}^{'} $, from our previous analysis in Section \ref{sec-4}, we have that idea-II will not be driven out either. Therefore the two ideas coexist. An example is illustrated in figure \ref{fig_10c} where the spreading rate and influential factor of idea-I are $0.15$ and $0.5$ respectively, and the spreading rate of idea-II remains as $0.25$. We see that the two ideas steadily coexist.

\begin{figure}
\begin{center}
\includegraphics[width=70mm,height=50mm]{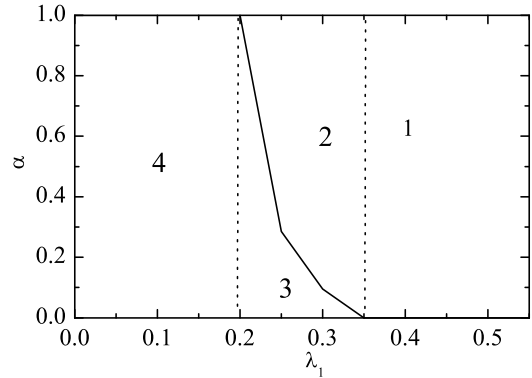} 
\end{center}
\caption{\label{fig_11}Schematic phase diagram of two ideas with exclusive and non-exclusive influences respectively in an infinite scale-free network with exponent value $r=3$ and the minimum nodal degree 2. The spreading rate of idea-II is fixed to be 0.25. In regions 1 and 2, idea-II is driven out. In region 3, there are multiple endemic states; while in region 4, the ideas always coexist.}
\end{figure}

Figure \ref{fig_11} demonstrates the different value regions of $\left(\lambda _{1},~\alpha \right)$ leading to different final outcomes and the phase transition in between. The calculations are based on an infinite scale-free network with exponent value $r=3$ and the minimum nodal degree $2$. The spreading rate of idea-II is fixed to be $0.25$. In region $1$, idea-I has a relatively high spreading rate and hence can always exclude idea-II regardless its influential factor value. When the spreading rate of idea-I gets lower, the influential factor starts to play a critical role in determining the final outcome: if $\alpha $ is high enough to be in region $2$, idea-I always survives and excludes idea-II; otherwise, in region $3$ the two ideas may have different endemic states depending on their initial densities. When the spreading rate of idea-I is further lowered and enters region $4$, the two ideas steadily coexist. Such conclusions reveal that when the two ideas are of comparable spreading rates, the neighborhood influence plays a critical role in determining the stationary state.

\subsection{ \label{subsec-5-2}Numerical results and discussions}
\begin{figure}
\begin{center}
\subfigure[]{\label{fig_12a}
\includegraphics[width=70mm,height=50mm]{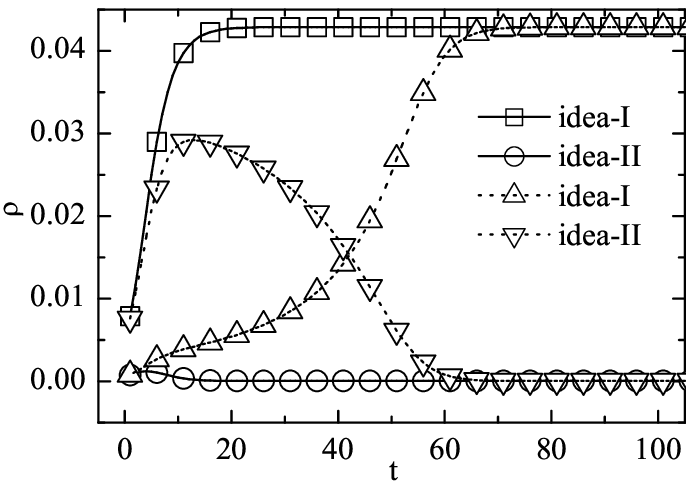}}
\subfigure[]{\label{fig_12b}
\includegraphics[width=70mm,height=50mm]{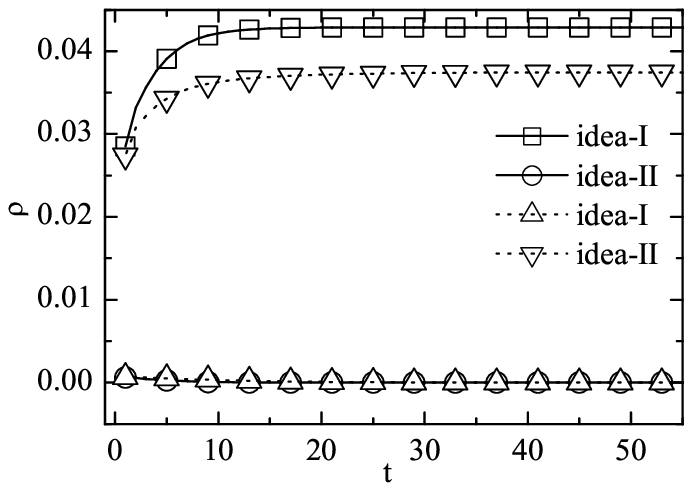}}
\end{center}
\caption{\label{fig_12}Time evolution of the densities of competing ideas with exclusive and nonexclusive influences respectively. The simulation is on top of a $10,000$-node random scale-free network with exponent $r=3$ and minimum degree $2$.  The spreading rates of idea-I and idea-II are fixed to be $\lambda_{1}=0.26$ and $\lambda_{2}=0.25$ respectively. In (a), the influential factor of idea-I is $\alpha=0.25$, while in (b), it is $\alpha=0.1$. The solid and dotted lines represents two separate sets of simulations with different initial densities. Results are averaged over at least 100 realizations.}
\end{figure}
\begin{figure}
\begin{center}
\includegraphics[width=70mm,height=50mm]{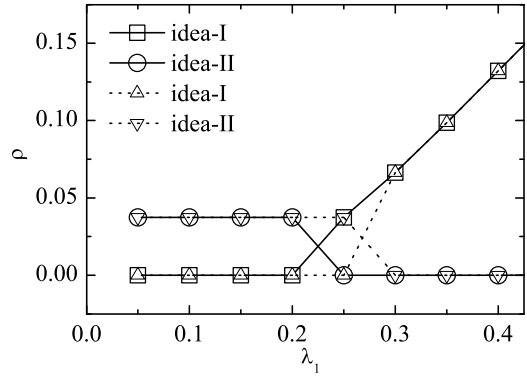} 
\end{center}
\caption{\label{fig_13}Stationary densities of competing ideas where the product of the influential factor and spreading rate of idea-I is of a fixed value at $0.025$. The simulation results are on top of a $10,000$-node scale-free network with exponent $r=3$ and the minimum nodal degree $2$. The spreading rate of idea-II is $\lambda _{2} = 0.25$. The solid and dotted lines represent two different sets of simulations with different initial densities $\rho _{1} (0)=0.5$, $\rho _{2} (0)=0.01$ (solid lines) and $\rho _{1} (0)=0.01$, $\rho _{2} (0)=0.5$ (dotted lines), respectively. Presented results are averaged over at least $100$ realizations.}
\end{figure}

The simulation results presented in this section are mainly for illustrating the effects of the different values of $\alpha$ when $\lambda _{1,c}^{'} \le \lambda _{1} <\lambda _{1,c} $. The same random scale-free network as that in Section~\ref{sec-4} is adopted. In figure \ref{fig_12}, the spreading rates of idea-I and idea-II are $\lambda _{1} =0.26$ and $\lambda _{2} =0.25$ respectively. The values of influential factor are $\alpha =0.25$ in figure~\ref{fig_12a} and $\alpha =0.1$ in figure~\ref{fig_12b}. Figure~\ref{fig_12a} shows that when $\alpha $ is high enough, idea-II is always driven out even when it has a relatively high initial density. However, when the influential factor is of a low value, e.g., $\alpha =0.1$, figure~\ref{fig_12b} shows that the final densities of the competing ideas may vary with their initial densities (Note that when idea-I has a low initial density, it may be driven to a very low final density but would still coexist with idea-II.). These observations match our analytical results that the two ideas may have multiple stationary endemic states. 

To illustrate the trade-off between increasing spreading rate and weakening the neighborhood influence imposed by the competitor when subject to limited resources, figure~\ref{fig_13} illustrates the stationary densities of the competing ideas where the product of the spreading rate and the influential factor of idea-I is fixed at $0.025$. The spreading rate of idea-II is $0.25$. Similar to the conclusions in Section~\ref{sec-3} on two competing ideas both with non-exclusive influences, it turns out to be a more effective strategy to increase spreading rate (i.e., increasing $\lambda_{1}$) rather than weakening the influence of the competitor (i.e., increasing $\alpha $). The stationary density of idea-I steadily increases with its spreading rate, and such increase can be rapid when the spreading rate of idea-I is high enough.

\section{\label{sec-6}Concluding remarks}
While competing agents with no neighborhood influence steadily coexist in scale-free networks, competing ideas with neighborhood influences have much richer dynamics. We considered different cases where the two competing ideas may have exclusive or non-exclusive influences. The study results help better understand the rise and fall of competing ideas in social systems and human society. 

It is probably least surprising that two ideas both with non-exclusive influences always co-exist. What may be more significant, however, is that these ideas may have multiple co-existences, where the stationary densities of the ideas are largely determined by their respective initial densities unless they have significantly different spreading rates. A novel idea, as a ``newcomer'' with a virtually zero initial density, may easily be suppressed to a low stationary density (i.e., a small size of population accepting the idea) unless, or until, it acquires a much higher spreading rate than that of the old, established idea. This remains as the case even if the newcomer can strongly suppress the spreading of the old idea. Such observations may help explain why it is usually difficult for a new idea to get wide acceptance. Acquiring popularity is the most effective, and in many cases may be the only way for the new idea to prevail. 
 
It is interesting that when both ideas are with exclusive influences, they can never stably co-exist: zero-tolerant extremists with different beliefs indeed can hardly live with each other in the same social system. And once again, unless the newcomer acquires a much higher spreading rate than that of the established one, an invading extremism idea may be easily driven out. Penetrating into an area under extensive control by extremism is a challenge for any new idea. The new idea has to have strong enough popularity in order to survive.
 
When extremism meets non-extremism, it is not a surprise that the extremism idea has a chance to eliminate its competitor altogether. This, however, is guaranteed to happen only when the non-extremism idea has a significantly lower spreading rate or a comparable spreading rate yet a much lower initial density. As long as the spreading rate of the non-extremism idea is high enough, it will survive or even prevail despite the fierce suppression from its extremism competitor. In fact, it is always a more effective strategy to increase the transmissibility of the idea rather than weakening the neighborhood influence of the competitor. Tolerance to the influences of competitors, if it helps focus on increasing transmissibility, may finally pay off. Such observations may help explain how tolerance started in the first place and why tolerance itself has become popular and ``politically correct''.
 
In our study, we have adopted the assumptions that the chance of accepting an idea depends on the presence of this idea in the neighborhood area rather than the number of neighbors accepting it; and co-infection is allowed on any individual. As pointed out in Section~\ref{sec-2}, the main conclusions presented in this paper shall basically still hold if we change the assumptions to allow each infectious neighbor has an independent chance of propagating the idea or to not allow co-infection. It is, however, not clear yet how degree correlations and community/clustering structures \cite{ref-28,ref-33,ref-34} can affect the final states of the competing ideas. Also, our study has been based on the random scale-free network model. Though simulation results did show that all the conclusions hold in real-life scale-free social networks (omitted in the paper due to length limit), e.g., the co-author network \cite{ref-35}, it remains largely unknown the dynamics of competing ideas in non-scale-free social networks. For the cases where the complex network itself also evolves, the dynamics of competing ideas are expected to be even richer. Such topics will be of our future research interest. 

In \cite{ref-36}, it was pointed out that in some social systems, there may exist zealots who never change their ideas. A zealot is different from an individual with exclusive influence: the former one sticks to an idea, whereas the latter one prohibits his/her neighbors from accepting any other idea. It may be very interesting and with significant importance to investigate the dynamics of a complex system with extremists, zealots, as well as regular individuals with reasonable tolerance. This will also be of our future research interest.

\ack{This work was partially supported by Ministry of Education, Singapore, under contract RG27/09.}


\end{document}